\documentclass[12pt,preprint]{aastex}

\begin{document}
\shortauthors{Bono et al.}
\shorttitle{The Cepheids in the maser-host galaxy NGC 4258}

\title{Cepheids in external galaxies. I. The maser-host galaxy NGC~4258 and the metallicity
dependence of $P$-$L$ and $P$-$W$ relations}

\author{G. Bono\altaffilmark{1,2}, 
F. Caputo\altaffilmark{1}, 
G. Fiorentino\altaffilmark{3},
M. Marconi\altaffilmark{4}, 
I. Musella\altaffilmark{4}
}

\altaffiltext{1}{INAF $-$ Osservatorio Astronomico di Roma, Via Frascati
33, 00040, Monte Porzio Catone, Italy}

\altaffiltext{2}{European Southern Observatory, Karl-Schwarzschild-Str. 2, 
85748 Garching bei Munchen, Germany} 

\altaffiltext{3}{Kapteyn Astronomical Institute, University of Groningen, Postbus 800, 9700 AV Groningen, the Netherlands}

\altaffiltext{4}{INAF $-$ Osservatorio Astronomico Di Capodimonte, Via Moiariello 16, 131 Napoli, Italy.}


\begin{abstract}

We perform a detailed analysis of Cepheids in NGC~4258,
Magellanic Clouds and Milky Way in order to verify the reliability
of the theoretical scenario based on a large set of nonlinear
convective pulsation models.
We derive Wesenheit functions from the synthetic $BVI$ magnitudes of the pulsators 
and we show that the sign and the extent
of the metallicity effect on the predicted Period-Wesenheit ($P$-$W$) relations 
change according to the adopted passbands.
These $P$-$W$ relations are applied to measured $BVI$ magnitudes of NGC~4258, 
Magellanic and Galactic Cepheids available in the literature.
We find that Magellanic and Galactic Cepheids agree
with the metallicity dependence of the predicted $P$-$W$ relations. 
Concerning the NGC~4258 Cepheids, the results strongly depend on 
the adopted metallicity gradient across the galactic disc. 
The most recent nebular oxygen abundances support a shallower 
gradient and provide a metallicity dependence that agrees well 
with current pulsation predictions. Moreover, the comparison of 
Cepheid distances based on $VI$ magnitudes with distance estimates 
based on the revised TRGB method for external galaxies, on the $HST$
trigonometric parallaxes for Galactic Cepheids, and on eclipsing
binaries in the Magellanic Clouds seems to favor the metallicity
correction predicted by pulsation models.
The sign and the extent of the metallicity
dependence of the Period-Wesenheit and of the Period-Luminosity
relations change according to the adopted passbands. Therefore,
distances based on different methods and/or bands should not be
averaged. The use of extragalactic Cepheids to constrain the
metallicity effect requires new accurate and extensive nebular
oxygen measurements.
\end{abstract}


%


\section{Introduction}

The Period-Luminosity ($P$-$L$) relation of Classical Cepheids is a
yardstick in several astrophysical and cosmological problems. The
Cepheid distances to external galaxies rely on fiducial $P$-$L$
relations based on Large Magellanic Cloud (LMC) variables and these
distance determinations are used to calibrate secondary distance
indicators, and in turn to estimate the Hubble constant $H_0$.
However, a general consensus on the ``universality'' of the $P$-$L$
relations, and in particular on their dependence on the Cepheid
chemical composition has not been reached yet.

On the theoretical side, it is worth mentioning that the nonlinear
convective pulsation models computed by our group (Fiorentino et al.
2007, hereinafter [F07]; Caputo 2008, and references therein) show that the
metallicity effect on the predicted $P$-$L$ relations depends on the
adopted photometric band. The synthetic linear $P$-$L$ relations,
for an increase in the global metal content from $Z$=0.004 to 0.02,
become on average shallower, with the slope of the optical
$P$-$L_B$, $P$-$L_V$ and $P$-$L_I$ relations decreasing from
$\sim$29\%, to 15\% and to 8\%, respectively. The same change in
metallicity causes no significant effect on the near-infrared
$P$-$L$ relations. Moreover, quoted predictions also indicate that
the metal-rich pulsators with periods longer than five days present
{\it fainter} optical magnitudes than the metal-poor ones. The
extent depends once again on the adopted passband. At even larger
metal abundances $Z$=0.03-0.04, the pulsation models suggest that
the helium content $Y$ also affects the Cepheid properties at
periods longer than about ten days. It was also suggested
(Fiorentino et al. 2002, hereinafter [F02]; Marconi, Musella \& 
Fiorentino 2005, hereinafter [M05]) that the metallicity effect 
on Cepheid distances {\it based
on $V$ and $I$ magnitudes}, is not linear over the entire
metallicity range $Z$=0.004-0.04, but presents a sort of
``turnover'' at roughly solar chemical composition. As a whole, the
use of LMC-calibrated $P$-$L_V$ and $P$-$L_I$ relations to provide
distance estimates with an intrinsic error of $\pm$0.10 mag is fully
justified for Cepheids with $P\le$ 10 days and/or helium-to-hydrogen
enrichment ratio $\Delta Y/\Delta Z\le$ 2.0 (see e.g. F02 and M05
for more details). On the other hand, the average correction for
Cepheids with $P>$ 10 days, high metal abundances ($Z\ge$ 0.03) and
$\Delta Y/\Delta Z\ge$ 3.0 is larger than 0.1 mag. As a consequence,
Cepheids with $P\ge$ 20 days and oxygen abundance\footnote{According
to the conventional logarithmic scale of stellar chemical
abundances, for two different elements $x_i$ and $x_j$ one has
[$x_i/x_j$]=log$(x_i/x_j)_*-$log$(x_i/x_j)_{\odot}$.} [O/H]$\ge$
0.2, as measured in several spiral galaxies observed by the $HST$
Key Projects (Freedman et al. 1994; Saha et al. 1994), the average
metallicity correction varies from about $-$0.2 mag to $\sim$ +0.25
mag as the adopted  $\Delta Y/\Delta Z$ ratio varies from 2 to 3.5.

On the observational side, independent investigations suggest either
a negligible metallicity effect or that Galactic (metal-rich)
Cepheids are somehow {\it brighter}, at fixed period, than LMC
(metal-poor) variables (Sasselov et al. 1997; Kennicutt et al. 1998,
2003; Kanbur et al. 2003; Tammann et al. 2003; Sandage et al. 2004;
Storm et al. 2004; Groenewegen et al. 2004; Sakai et al. 2004; Ngeow
\& Kanbur 2004; Pietrzynski et al. 2007). In the latter case, the
empirical metallicity dependence of the Cepheid true distance
modulus $\mu_0$, as usually described by the parameter
$\gamma=\delta\mu_0/\delta$log$Z$ where $\delta\mu_0$ is the extent
of the metallicity correction and
$\delta$log$Z$=log$Z_{LMC}-$log$Z_{Ceph}$, spans a large range of
{\it negative} values up to $\gamma=-$0.4~mag~dex$^{-1}$, with an
average value of approximately $-$0.25~mag~dex$^{-1}$ (Sakai et al.
2004 and references therein). However, spectroscopic
iron-to-hydrogen [Fe/H] measurements of Galactic Cepheids
(Romaniello et al. 2005) indicate that the visual $P$-$L_V$ relation
depends on the metal content, but exclude that the metallicity
correction follows the linear relation based on the quoted negative
empirical $\gamma$-value.
The nonlinear behavior suggested by the pulsation models accounts
quite well for the observed trend.

More recently, Macri et al. (2006, hereinafter [M06]) have presented multiband
$BVI$ observations of a large Cepheid sample in two fields of the
galaxy NGC~4258 with different mean chemical compositions ($\Delta$
[O/H]$\sim$0.5 dex), and derived a metallicity effect of
$\gamma=-$0.29~mag~dex$^{-1}$, also excluding any significant
variation in the slope of the $P$-$L$ relations as a function of the
Cepheid metal abundance.
 Their findings agree quite well with the results of a previous
investigation by Kennicutt et al. (1998) who found
$\gamma=-$0.27~mag~dex$^{-1}$ using a sizable sample of Cepheids in
two fields of M~101 with a difference in mean oxygen abundance of
0.7 dex.

A vanishing metallicity effect between Galactic and Magellanic Cepheids 
was also found by Fouque et al. (2007). They collected a sample of  
59 calibrating Galactic Cepheids with distances based on robust indicators:
HST and Hipparcos trigonometric parallaxes, infrared surface brightness, 
Interferometric Baade-Wesselink methods and cluster main-sequence fitting.  
By comparing the slopes of Galactic optical-NIR PL and Wesenheit relations  
with LMC slopes provided by OGLE (Udalski et al. 1999a) and by Persson et al.
(2004) they find no significant difference. 
Accurate trigonometric parallaxes for ten Galactic Cepheids have been
provided by Benedict et al. (2007) using the Fine Guide Sensor available
on board the HST. They estimated new optical and NIR PL relations and
they found that their slopes are very similar to the slopes of LMC
Cepheids.\\
However, Mottini et al. (2008) analyzed a sizable sample of high-resolution, 
high signal-to-noise spectra collected with FEROS@1.5m ESO telescope for 
Galactic (32) and with UVES@VLT for Magellanic (14 in SMC and 22 in LMC) 
Cepheids. They found, using individual iron measurements and the same distances
adopted by Fouque et al., that the slope of the $V-$band PL relation 
does depend on the metal abundance with a confidence level larger than 90\%.

In order to overcome current controversy between theory and
observations, we undertook a homogeneous analysis of the NGC~4258
Cepheids and a detailed comparison of pulsation predictions with
Magellanic Cloud and Galactic Cepheids. In \S~2 we present predicted
$P$-$L$ relations, while in \S~3 we describe the results of the
comparison between theory and observations. The correlation between
the Cepheid metallicity and the NGC~4258 oxygen abundance gradient
is addressed in \S~4 and the conclusions close the paper.

\section{Pulsation models}

The fiducial $P$-$L$ relations adopted by M06 are based on
unreddened $B_0,V_0,I_0$ magnitudes of the LMC Cepheids observed by
the OGLE II project (Udalski et al. 1999a, hereinafter [U99]) and updated on the
OGLE Web site\footnote{http://www.astrouw.edu.pl/$\sim$ogle}. They are

\begin{equation}B_0=17.368-2.439\log P\end{equation}
\begin{equation}V_0=17.066-2.779\log P\end{equation}
\begin{equation}I_0=16.594-2.979\log P\end{equation}
\noindent where $P$ is the pulsation period in days. These relations
are used to form intrinsic Period-Color ($P$-$C$) relations which
are compared with the measured colors to determine the $E(B-V)$,
$E(V-I)$, and $E(B-I)$ reddening values for individual Cepheids in
NGC~4258. Then, the absolute LMC-relative distance modulus
$\delta\mu_0$ of each variable is derived by averaging the three
values\footnote{The $A_I/E(B-I)$=2.38 value given by M06 is a typo.}
$$\delta\mu_{0,VI}=\delta\mu_I-1.45E(V-I)$$
$$\delta\mu_{0,BI}=\delta\mu_I-0.82E(B-I)$$
$$\delta\mu_{0,BVI}=\delta\mu_I-1.94E(B-V)$$
\noindent where $\delta\mu_I$ is the difference between the observed
$I$ magnitude and the $I_0(P)$ value from equation (3), while the
$A_I/E(B-I)$, $A_I/E(V-I)$, and $A_I/E(B-V)$ ratios are based on the
$A_\lambda/E(B-V)$ values from Table 6 in Schlegel et al. (1998) for
$A_V/E(B-V)$=3.1 and the Cardelli et al. (1989) extinction law.
However, since equations (1)-(3) were derived by adopting
$A_B/E(B-V)$=4.32, $A_V/E(B-V)$=3.24 and $A_I/E(B-V)$=1.96 (see
U99), throughout this paper we adopt

\begin{equation}\delta\mu_{0,VI}=\delta\mu_I-1.53E(V-I)\end{equation}
\begin{equation}\delta\mu_{0,BI}=\delta\mu_I-0.83E(B-I)\end{equation}
\begin{equation}\delta\mu_{0,BVI}=\delta\mu_I-1.96E(B-V),\end{equation}
\noindent together with

\begin{equation}\delta\mu_{0,BV}=\delta\mu_V-3.24E(B-V)\end{equation}
\noindent where $\delta\mu_V$ is the difference between the observed
$V$ magnitude and the $V_0(P)$ value from equation (2).

The quoted approach is equivalent to the classical method of
distance determinations based on the reddening free Wesenheit
functions. Therefore, we use the computed periods and
intensity-averaged $M_B,M_V,M_I$ magnitudes of our fundamental
pulsation models with $Z$=0.004 to 0.04, listed  in Table 1, to
derive the predicted Period-Wesenheit ($P$-$W)$ relations based on
equations (4)-(7), i.e.
$$WVI=M_I-1.53(M_V-M_I)$$
$$WBI=M_I-0.83(M_B-M_I)$$
$$WBVI=M_I-1.96(M_B-M_V)$$
$$WBV=M_V-3.24(M_V-M_B)$$

In our earlier pulsation models, the adopted luminosity for a given
mass and chemical composition was fixed according to Mass-Luminosity
($ML)$ relations based on canonical (``can'') evolutionary
computations (Castellani, Chieffi \& Straniero 1992; Bono et al.
2000; Girardi et al. 2000). Afterwards, additional models have been
computed with higher luminosity levels (``over'') in order to
account for a mild convective core overshooting and/or mass loss
before or during the Cepheid phase. 
The overluminous models were constructed by adopting for the 
chemical compositions representative of Galactic and Magellanic 
Cepheids the same abundances (helium, metal) and mass values  
adopted for canonical Cepheid models. This approach allowed us 
(Bono, Castellani, \& Marconi 2000; Caputo et al. 2005) to
constrain the impact that the mass-luminosity relation has on
pulsation observables.
The quoted assumptions concerning the adopted chemical 
compositions are supported by theory and observations. Pulsation 
models constructed by adopting supersolar iron abundance and helium 
enhanced compositions ($\frac{\Delta Y}{\Delta Z}=4$) are pulsationally 
unstable (Fiorentino et al. 2002). Moreover, empirical 
evidence based on He abundance of Planetary Nebulae suggest a very 
shallow gradient across the Galactic disk (Stanghellini et al. 2006). 
Furthermore, chemical evolution models for both the inner and the 
outer disk indicate similar helium gradients (Hou et al. 2000). 
Note that current estimates of the helium-to-metal enrichment ratio,
$\frac{\Delta Y}{\Delta Z}$, are still affected by large uncertainties.
In a recent detailed investigation Casagrande et al. (2007) found, using
nearby field K-type dwarf stars, $\frac{\Delta Y}{\Delta Z}=2.2\pm1.1$.
The observational scenario is also complicated by the fact that we still
lack firm empirical constraints on the linearity of the
$\frac{\Delta Y}{\Delta Z}$ relation, when moving from the metal-poor
to the metal-rich regime, and on the universality of this relation
(Peimbert et al. 2003; Tammann et al. 2008). To account for these uncertainties
we constructed sets of Cepheid models by adopting, at fixed metal content,
different helium abundances, therefore, the intrinsic error on the zero-point
of predicted PL and PW relations include this effect (Fiorentino et al. 2007).

The theoretical Wesenheit functions of each pulsator depend on the 
adopted luminosity, therefore,  to avoid any assumption on the 
$ML$ relation, we calculated for each model the difference 
log$L/L_c$ between the adopted luminosity and the canonical value 
provided by the Bono et al.  (2000) relation

\begin{equation}
\log L_c=0.90+3.35\log M_e+1.36\log Y-0.34\log Z
\end{equation}

where mass and luminosity values are in solar units. Then,
by a linear interpolation through all the fundamental models with
period $P\sim$ 4-80 days, $Z$=0.004-0.04, without distinction of the
helium content at fixed $Z$, we derive the linear Period-Wesenheit
($P$-$W$) relations listed in Table 2.

According to these predicted $P$-$W$ relations, one can determine
the true distance modulus $\mu_0$ of individual Cepheids with known
metal content, once the log$L/L_c$ ratio is fixed. In this context,
it is worth mentioning that the occurrence either of a mild
convective core overshooting during hydrogen burning phases or
mass-loss before and/or during the pulsation phases yields positive
log$L/L_c$ values. As a consequence, the $P$-$W$ relations for
$L$=$L_c$ provide the {\it maximum} value of the Cepheid distance.
Moreover, we draw attention on the evidence that the metallicity
effect on the predicted $P$-$W$ relations depends on the adopted
Wesenheit function (see also Caputo, Marconi \& Musella 2000; F02;
F07). In particular, the metallicity dependence of the $P$-$WVI$
relation is weak and shows the opposite sign when compared with the
other optical $P$-$W$ relations. This is warning against the average
of the various $\mu_0$ values based on different $P$-$W$ relations
and, at the same time, provides a plain method to estimate the
Cepheid metal content (see below).

By assuming that the pulsation models listed in Table 1 are actual
Cepheids located at the same distance and with the same reddening,
but with different chemical abundances, we can derive from equations
(1)-(3) their LMC-relative apparent distance moduli $\mu_B$, $\mu_V$
and $\mu_I$. Then, by adopting\footnote{The comparison of observed
colors with the intrinsic $P$-$C$ relations given by equations
(1)-(3) to estimate the reddening values is equivalent to take the
differences between two LMC-relative apparent distance moduli and
indeed
$E(B-V)=(B-V)-(B_0-V_0)=(B-B_0)-(V-V_0)=\delta\mu_B-\delta\mu_V$.}
$\mu_B-\mu_V$=$E(B-V)$, $\mu_B-\mu_I$=$E(B-I)$ and
$\mu_V-\mu_I$=$E(V-I)$, we use equations (4)-(7) to determine the
four LMC-relative intrinsic values
 $\delta\mu_{0,VI}$, $\delta\mu_{0,BI}$, $\delta\mu_{0,BVI}$,
and $\delta\mu_{0,BV}$.

\clearpage
\begin{figure}
\begin{center}
\includegraphics[width=8cm]{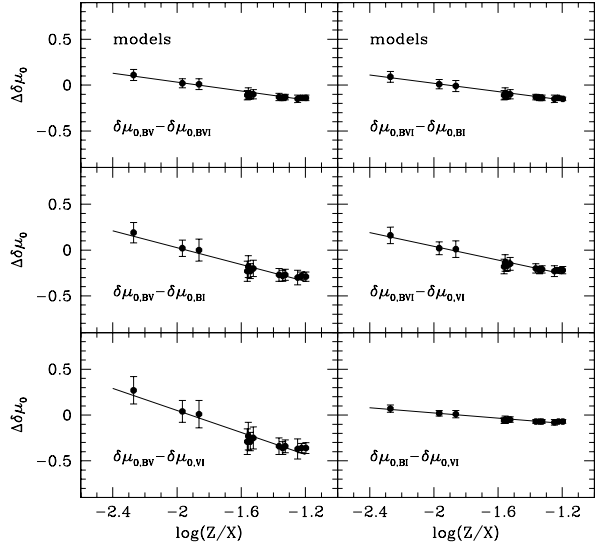}
\caption{Internal differences among LMC-relative distance moduli for
fundamental pulsators with log$P$=0.4-1.9 versus the chemical
composition, as listed in Table 4. The lines display the least-squares
fit to the data.}
\end{center}
\end{figure}
\clearpage

By averaging the results over the entire period range ($P\sim$ 4-80
days), without selections between short and long period pulsators,
we get the LMC-relative $\delta\mu_0$ values at $L=L_c$ listed in
Table 3. As already suggested by the predicted $P$-$W$ relations
given in Table 2, we find that the metallicity effect is {\it not
constant} among the different approaches to estimate the
LMC-relative distance moduli. On average, we derive
$\gamma(\delta\mu_{0,BV})\sim-$0.59~mag~dex$^{-1}$,
$\gamma(\delta\mu_{0,BI})\sim-$0.12~mag~dex$^{-1}$, and
$\gamma(\delta\mu_{0,BVI})\sim-$0.35~mag~dex$^{-1}$, whereas the
metallicity dependence of $\delta\mu_{0,VI}$ is smaller and seems to
depend on the adopted metallicity range. We find
$\gamma(\delta\mu_{0,VI})\sim$ +0.11~mag~dex$^{-1}$ for $Z\le$ 0.02
and $\sim-$0.15 for $Z\ge$ 0.02, while over the entire metallicity
range $Z$=0.004-0.04, we find $\gamma(\delta\mu_{0,VI})\sim
+$0.03~mag~dex$^{-1}$.

In summary, the theoretical results suggest that, if the LMC-based
$PL_B$, $PL_V$ and $PL_I$ relations are used to get the distance to
Cepheids with metal content significantly different from the LMC
abundance, then the values of the various $\delta\mu_0$ formulations
{\it should not be averaged, but individually considered in order to
keep the information provided by their different metal dependence}.
 Note that if the $\delta\mu_{0,VI}$, $\delta\mu_{0,BI}$, and
$\delta\mu_{0,BVI}$ values were averaged to a mean value
$\langle\delta\mu_0\rangle$, the ensuing mean metallicity effect
would be
$\gamma(\langle\delta\mu_0\rangle)\sim-$0.15~mag~dex$^{-1}$. This
value {\it cannot} be used to correct distance estimates based on
$VI$ magnitudes since it might introduce a systematic error up to
$\approx$ 0.2 mag according to the metallicity range covered by the
Cepheids.

A further relevant result of the present study is the predicted
metallicity effect on the {\it internal differences} among the different  
LMC-relative distance moduli. In particular, these differences revealed 
to be almost independent of the adopted $ML$ relation.
Data plotted in Fig.~1 and listed in Table 4 show that all the differences
depend on the pulsator chemical composition. The most metal-sensitive
are the $\Delta\delta\mu_{0,BV-VI}$, the $\Delta\delta\mu_{0,BV-BI}$ and
the $\Delta\delta\mu_{0,BVI-VI}$. We have already discussed in F07 that
these differences provide a robust method to estimate the Cepheid metal
content, since they are independent of both distance and reddening.

\section{Observed Cepheids}
\subsection{Magellanic and Galactic variables}
\clearpage
\begin{figure}
\begin{center}
\includegraphics[width=8cm]{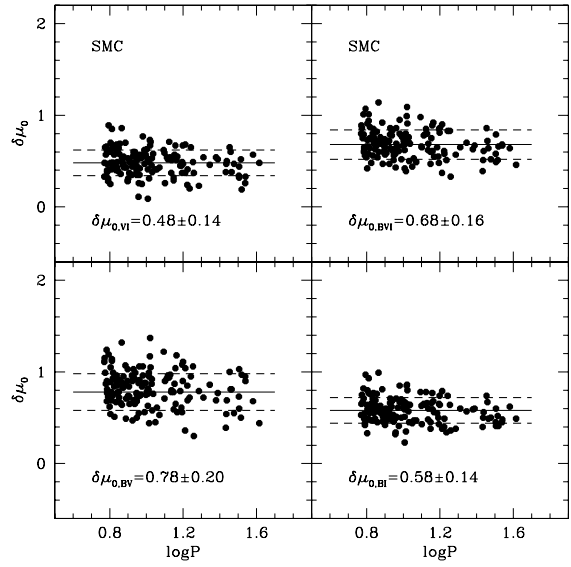}
\caption{LMC-relative distance moduli for OGLE II
SMC fundamental Cepheids versus the period. Solid and dashed lines display
the average values and the standard deviations, respectively.}
\end{center}
\end{figure}

\begin{figure}
\begin{center}
\includegraphics[width=8cm]{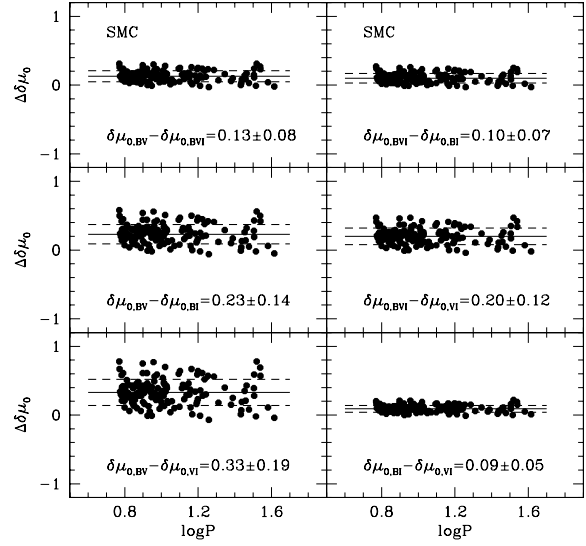}
\caption{Differences among LMC-relative distance moduli for OGLE II
SMC fundamental Cepheids versus the period. Solid and dashed lines display
the average values and the standard deviations, respectively.}
\end{center}
\end{figure}
\clearpage

We apply the same procedure described above to the SMC Cepheids
collected by the
OGLE-II microlensing survey (Udalski et al. 1999b). For the sake of
uniformity with the NGC~4258 variables (see later), we use
fundamental pulsators with $P\ge$ 6 days ($\sim$ 200 variables). 
We derive the mean
$\delta\mu_0$ values plotted in Fig. 2 and summarized in Table 5,
where the results for the LMC Cepheids (U99) are also given to
validate the current approach.

Even though data plotted in Fig.~2 present a large scatter, the four
LMC-relative $\delta\mu_0$ formulations do provide different
results. In particular, the $\delta\mu_{0,VI}$ gives the shortest
LMC-relative distance modulus, in agreement with the theoretical
predictions. By using the predicted metallicity effects given in the
last row of Table 5 and the Cepheid spectroscopic measurements
[Fe/H]$_{LMC}=-0.35\pm0.15$ dex and [Fe/H]$_{SMC}=-0.70\pm0.15$ dex
(see Luck et al. 1998; Romaniello et al. 2005) and by assuming
$\Delta[Z/X]$=$\Delta$[Fe/H], we find that the measured
$\delta\mu_{0,BV}$, $\delta\mu_{0,BI}$, and $\delta\mu_{0,BVI}$
values should be {\it decreased} by $\sim$ 0.21, 0.04 and 0.12 mag,
whereas the $\delta\mu_{0,VI}$ value should be {\it increased} by
$\sim$ 0.04 mag. Eventually, the discrepancy among the four
$\delta\mu_0$ values is mitigated and the metallicity-corrected
results yield that the LMC-relative distance modulus of the SMC
Cepheids is $\sim$ 0.55 mag, in close agreement with the difference
of 0.50 mag determined from eclipsing binaries in the SMC (Hilditch,
Howarth \& Harries 2005) and in the LMC (Guinan, Ribas \&
Fitzpatrick 2004).

The differences $\Delta\delta\mu_0$ among the four
$\delta\mu_0$ values  are summarized in Table 6 and plotted in
Fig.~3. The observed variations between SMC and LMC Cepheids agree
quite well with pulsation predictions for $\Delta$[Z/X]=$\Delta$[Fe/H]=$-$0.35,
as listed in the last row of Table 6.
In addition, the straight comparison between the observed
$\Delta\delta\mu_0$ differences and the predicted values listed in
Table 4 gives log$(Z/X)_{LMC}\sim -$1.92 and log$(Z/X)_{SMC}\sim
-$2.42, which are consistent with the spectroscopic iron
measurements once we assume [Fe/H]=[$Z/X$] and $(Z/X)_{\odot}$=0.024
(Grevesse et al. 1996). We are aware that the solar chemical
composition is under revision and that the recent analysis by
Asplund et al. (2004) has decreased the solar chemical abundances 
by roughly a factor of two, yielding $(Z/X)_{\odot}$=0.0165.
However, we adopted the Grevesse et al. (1996) solar abundances, since
they are consistent with the model atmospheres (Castelli, Gratton \&
Kurucz 1997a,b) we use to transform theoretical predictions into the
observational plane. The revised abundances are still debated due to 
the inconsistency with helioseismic results (Bahcall et al. 2005; 
Guzik et al. 2005). Following the referee's suggestion, it is also 
worth noting that evolutionary and pulsation models are constructed 
by adopting the global metallicity ([M/H]) that
is a function of both iron and $\alpha$-element abundances. However,
evolutionary prescriptions by Salaris et al. (1993) to estimate the
global metallicity were derived using the old solar mixture and we
still lack a new relation based on the new solar abundances. 
Moreover, Salaris \& Weiss (1998) found 
that at solar and super-solar metallicities, the metals affects 
the evolution. This means that scaled-solar abundances cannot be 
used to replace the $\alpha$-enhanced ones of the same total 
metallicity. Oxygen is an $\alpha$-element and intermediate-mass 
stars with solar abundance typically present solar Oxygen abundances 
(Gratton et al. 2004). Empirical evidence indicates that field
LMC giants present a lower $\alpha$-enhancement when compared with
the Galactic ones  (Luck et al. 1998; Hill et al. 2000). However, 
larger samples are required to constrain the trend in the metal-poor 
regime (Hill 2004; Venn et al. 2004).
It is worth noting that the new solar abundances imply a change in 
the nebular Oxygen abundance of external galaxies (zero-point and 
effective temperature of the ionizing stars). To our knowledge we 
still lack nebular Oxygen abundances in external galaxies accounting 
for this effect.

\clearpage
\begin{figure}
\begin{center}
\includegraphics[width=8cm]{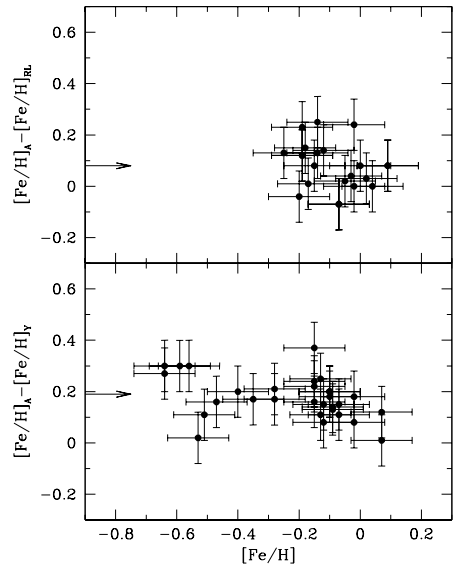}
\caption{Top -- Difference in iron abundance [Fe/H]$_A-$[Fe/H]$_{RL}$
versus [Fe/H]$_{RL}$ for Galactic Cepheids. The arrow shows the adopted
average difference. Bottom -- Same as the top, but for [Fe/H]$_A-$[Fe/H]$_Y$,
versus [Fe/H]$_Y$.
}
\end{center}
\end{figure}

\begin{figure}
\begin{center}
\includegraphics[width=8cm]{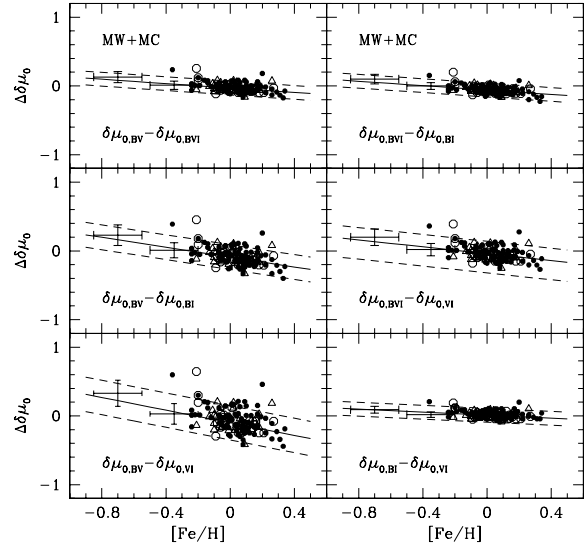}
\caption{Differences among LMC-relative distance moduli for Galactic
Cepheids with $P\ge$ 6 days versus the three different sets of [Fe/H]
measurements, [Fe/H]$_A$: solid circles, [Fe/H]$_Y$: open circles,
and [Fe/H]$_{RL}$: open triangles in the Andrievsky metallicity
scale. The mean values for SMC and LMC variables (crosses) are also
plotted. The solid lines are the least square fits to the data,
while the dashed lines display the dispersion around the fit.
See text for more details.}
\end{center}
\end{figure}

\clearpage
In order to verify whether this consistency between the pulsation
predictions and the Cepheid observed properties also holds up at
larger metal abundances, we use
the Milky Way variables with measured iron-to-hydrogen ratios. Given
the current discrepancy among abundance determinations by different
authors, we consider three different sets of measurements: the
[Fe/H]$_A$ values provided by Andrievsky and collaborators
(Andrievsky et al. 2002a,b,c; Andrievsky et al. 2004; Luck et al. 2003;
Kovtyukh, Wallerstein \& Andrievsky 2005; Luck, Kovtyukh \& Andrievsky 2006),
the [Fe/H]$_{RL}$ values by Romaniello et al. (2005) together with
Lemasle et al. (2007), and the [Fe/H]$_Y$ values by Yong et al. (2006)
together with Fry \& Carney (1997). Following Yong et al. (2006) the iron
abundances provided by Fry \& Carney were decreased by $-$0.11 dex.
The arrows plotted in Fig.~4 show that the [Fe/H]$_Y$ and the [Fe/H]$_{RL}$
were normalized to the Andrievsky metallicity scale by adding 0.19 and
0.08 dex, respectively.

We also use the $BVI$ magnitudes compiled by Berdnikov, Dambis \&
Vozyakova (2000). We select the variables with $P\ge$ 6 days,
although the inclusion of first overtone pulsators has no dramatic
effects on the {\it differences} among the $\delta\mu_0$ values. In
fact, adopting log$P_F$=log$P_{FO}+0.13$, one easily derives that
the offsets (first overtone minus fundamental) are $\sim-$0.08
($\Delta\delta\mu_{0,BV-VI}$), $-$0.06
($\Delta\delta\mu_{0,BV-BI}$), $-$0.03
($\Delta\delta\mu_{0,BV-BVI}$), $-$0.02
($\Delta\delta\mu_{0,BI-VI}$), $-$0.05
($\Delta\delta\mu_{0,BVI-VI}$), and $-$0.03~mag
($\Delta\delta\mu_{0,BVI-BI}$).

The results plotted in Fig.~5 show that Magellanic and Galactic
Cepheids follow reasonably well defined common relations over the
metallicity range [Fe/H]=$-$0.7 to +0.3, with the only exception of
CK~Pup, HQ~Car and TX~Cyg (at [Fe/H]=$-$0.36, $-$0.22 and +0.20,
respectively). Eventually, the linear regression through the
Magellanic and Galactic data yields the empirical
$\Delta\delta\mu_0$-[Fe/H] calibrations:
\begin{equation}\Delta\delta\mu_{0,BV-VI}=-0.10  (\pm0.14)-0.46(\pm0.08)[Fe/H]\;\;\;\;\sigma=0.25 \end{equation}
\begin{equation}\Delta\delta\mu_{0,BV-BI}=-0.09  (\pm0.10)-0.36(\pm0.08)[Fe/H]\;\;\;\;\sigma=0.18 \end{equation}
\begin{equation}\Delta\delta\mu_{0,BV-BVI}=-0.03 (\pm0.06)-0.16(\pm0.03)[Fe/H]\;\;\;\;\sigma=0.10 \end{equation}
\begin{equation}\Delta\delta\mu_{0,BI-VI}=+0.01  (\pm0.06)-0.11(\pm0.02)[Fe/H]\;\;\;\;\sigma=0.10 \end{equation}
\begin{equation}\Delta\delta\mu_{0,BVI-VI}=-0.04 (\pm0.09)-0.25(\pm0.05)[Fe/H]\;\;\;\;\sigma=0.18 \end{equation}
\begin{equation}\Delta\delta\mu_{0,BVI-BI}=-0.06 (\pm0.05)-0.16(\pm0.03)[Fe/H]\;\;\;\;\sigma=0.10 \end{equation}

where the error in parentheses is the error on the coefficients and
the sigma gives the sum in quadrature of the uncertainties affecting both
the zero-point and the slope of the fit. These relations are drawn as solid
lines in Fig.~5, while the dashed lines display the one $\sigma$ statistical
uncertainty.
It is worth emphasizing that the {\it observed} $\Delta\delta\mu_0$-[Fe/H]
relations based on Magellanic and Galactic Cepheids agree well with
the {\it theoretical} ones presented in Table 4, if we assume
[Fe/H]=[Z/X] and we adopt $(Z/X)_{\odot}$=0.024 (Grevesse et al.
1996), namely [Fe/H]=log$(Z/X$)+1.62. Note that an even better
agreement is found if we account for the measured overabundance of
$\alpha$-elements for subsolar [Fe/H] ratios, as determined by
spectroscopic measurements. (see, e.g., Fig. 18 in Yong et al.
2006). This issue will be discussed in a forthcoming paper.

\clearpage
\begin{figure}
\begin{center}
\includegraphics[width=8cm]{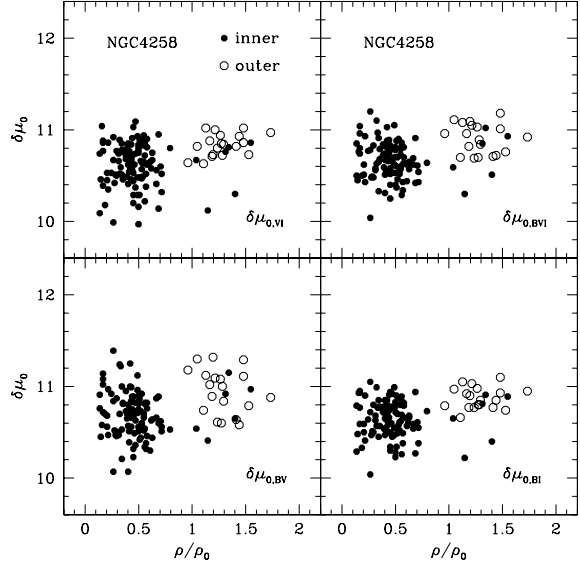}
\caption{LMC-relative distance moduli for NGC~4258 Cepheids with
$P\ge$ 6 days versus the deprojected radial distance $\rho(')$ normalized
to $\rho_0$=7'.92. Cepheids located either in the inner or in the outer
field are filled and open circles, respectively.}
\end{center}
\end{figure}

\begin{figure}
\begin{center}
\includegraphics[width=8cm]{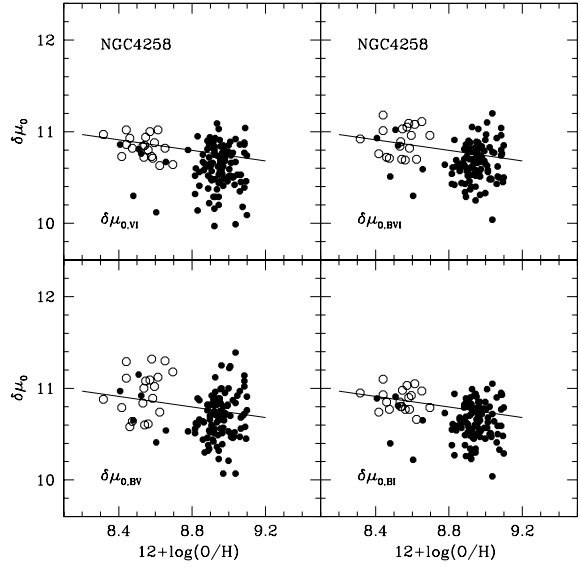}
\caption{Same as in Fig.~6, but with the LMC-relative distance moduli
versus the oxygen abundance based on the Za94 gradient. The solid line
shows the relation given by M06.}
\end{center}
\end{figure}

\begin{figure}
\begin{center}
\includegraphics[width=8cm]{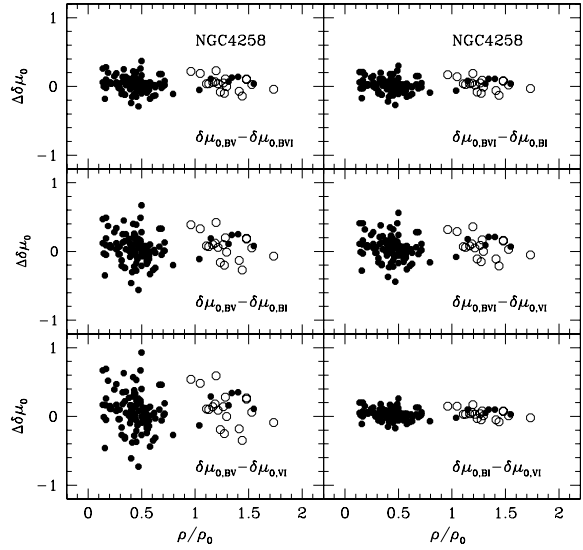}
\caption{Differences among LMC-relative distance moduli for NGC~4258
Cepheids with $P\ge$ 6 days versus the deprojected radial distance
$\rho/\rho_0$. Symbols are the same as in Fig.~6.}
\end{center}
\end{figure}
\clearpage

\subsection{NGC~4258 Cepheids}

The Cepheids observed in NGC~4258 belong to two different fields located at
different galactocentric distances and whose mean offsets in arcseconds from
the nucleus are $\approx -$150 (inner field) and $\approx +$400 (outer field)
in the East-West direction, while they are $\approx +$130 (inner field) and
$\approx -$400 (outer field) in the North-South direction.

We apply the same approach already adopted for Magellanic and Galactic
variables to the NGC~4258 Cepheids with a variability index
$L_V>$2 (``restricted'' sample in M06), $P\ge$ 6 days\footnote{The
$P_{min}$ adopted by M06 are 6 and 12 days for the Cepheids in the
outer and in the inner field, respectively. We adopt the same
period cuts for both fields in order to have homogeneous samples.}
and errors in the mean $BVI$ magnitudes less than 0.05 mag. The
derived LMC-relative distance moduli are plotted in Fig.~6 versus
the Cepheid deprojected galactocentric distance $\rho$(')
normalized to the isophotal radius $\rho_0$=7'.92, as
given\footnote{The $\rho/\rho_0$ values given by M06 adopt
$\rho_0$=7'.76. However, for consistency with the Zaritsky et al.
(1994, hereinafter [Z94]) abundance gradient, we normalized them to
$\rho_0$=7'.92.} by M06.

Data plotted in  Fig.~6 suggest a correlation
between the Cepheid distance modulus and its radial distance, with
the outer field Cepheids yielding larger distance moduli by about
0.2~mag with respect to those in the inner field. Obviously, such a
correlation turns into a chemical abundance dependence of the
distance modulus if a metallicity gradient is adopted for the Cepheids.
As a fact, using for each individual variable the Za94
relation based on oxygen abundance measurements of H~II regions
\begin{equation}12+\log(O/H)=8.97-0.49(\rho/\rho_0-0.4),\end{equation}
\noindent where $\rho_0$ is the isophotal radius equal to 7'.92, the
metallicity effect on the the four LMC-relative distance moduli
turns out to be consistent with the M06 value, i.e.
$\gamma\sim-$0.29~mag~dex$^{-1}$ (see Fig.~7).

\clearpage
\begin{figure}
\begin{center}
\includegraphics[width=8cm]{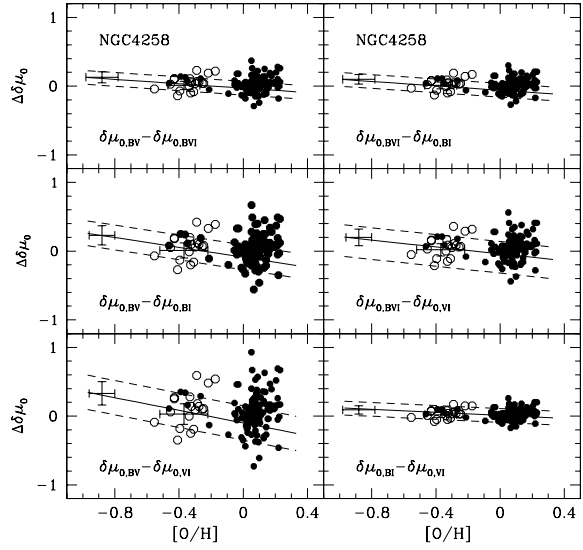}
\caption{Same as in Fig. 8, but with $\Delta\delta\mu_0$ versus the
oxygen-to-hydrogen ratio [O/H] based on the galactic gradient given
by Za94 and by adopting log(O/H)$_{\odot}=-$3.13. The crosses mark
the results for SMC and LMC variables. Solid and dashed
lines are the same as in Fig.~5, and have been plotted assuming 
[Fe/H]=[O/H].}
\end{center}
\end{figure}
\clearpage

However, we show in Fig.~8 that if the individual differences
$\Delta\delta\mu_0$ are taken into account, then no clear
radial dependence is found, with the Cepheids in both fields
yielding quite similar mean $\Delta\delta\mu_0$ values. Data 
plotted in Fig.~9 show the differences $\Delta\delta\mu_0$ for 
the NGC~4258 Cepheids versus the oxygen abundance 
[O/H]$_{Za94}$=log(O/H)$_{NGC4258}-$log(O/H)$_{\odot}$,
following equation (15) and by adopting the solar value 
log(O/H)$_{\odot}=-$3.13 (Grevesse et al. 1996). In order to 
make an easy comparison, in this figure we also plotted the mean 
$\Delta\delta\mu_0$ values for SMC and LMC Cepheids (see Table 6) 
for [O/H]=$-$0.88$\pm$0.08 dex and for $-$0.37$\pm$0.15 dex 
(Ferrarese et al. 2000). The least square fits to the data (solid 
lines) and their dispersions (dashed lines) showed in Fig.~5
are also plotted in Fig.~9 by assuming that the oxygen abundance is 
a very robust proxy of the iron abundance (i.e. [Fe/H]=[O/H]). Note that 
this assumption is fully justified by spectroscopic measurements which 
yield [O/Fe]=0$\pm$0.14 dex over the range [Fe/H]=$-$0.7 to +0.30 dex 
(see e.g. Luck et al. 2006).
Moreover, recent spectroscopic measurements based on high resolution,
high signal-to-noise spectra of 30 Galactic Cepheids (Lemasle et al. 2007)
indicate that Oxygen and other $\alpha$-elements present radial
gradients very similar to the iron gradient. 
This means that Oxygen is a good proxy of the iron content across the
Galactic disk.
Moreover and even more importantly, empirical evidence indicates that
Oxygen nebular abundances agree with absorption line abundances
(Hill 2004).

Although the oxygen abundance of the NGC~4258 Cepheids based on 
the Za94 oxygen abundance gradient, is within the range spanned 
by Magellanic and Milky Way Cepheids, the observed $\Delta\delta\mu_0$ 
values of several variables in the inner field deviate from the ``empirical''
$\Delta\delta\mu_0$-[O/H] relations provided by Magellanic and
Galactic variables. 
This evidence indicates that Cepheids in NGC~4258 might have a metal content
that is significantly lower than the oxygen abundance based on their radial
distance.

To make clear this feature, we select the inner field Cepheids with
$\rho/\rho_0<0.7$ (sample A) and the outer field Cepheids with
$\rho/\rho_0>1.0$ (sample B) for which the mean oxygen abundance
suggested by the Za94 gradient is [O/H]$_{Za94}$=+0.13$\pm$0.08 and
$-$0.37$\pm$0.09 dex, respectively. We show in Table~7 and Table~8
that the two samples have different LMC-relative distance moduli but
nearly identical mean $\Delta\delta\mu_0$ values, at odds with the
behavior of Magellanic and metal-rich Milky Way variables, as listed
in the same Table~8. In conclusion, the observed $\Delta\delta\mu_0$
values would suggest an average LMC-like oxygen abundance
[O/H]$\sim-$0.4 dex for all the NGC~4258 Cepheids. This result 
agrees with the Za94-based mean value of the outer field, whereas
for the inner field the oxygen content provided by the radial
distance appears to be 1.3 times larger than the solar value.

We do not found any reason to distrust this intriguing result, since
the selection criteria adopted by M06 are very robust, and indeed,
we only use objects with errors in mean $B,V,I$ magnitudes smaller
than 0.05 mag. The adopted $P_{min}$=6 days should avoid contamination
by first overtone Cepheids, although the effects of period uncertainties on
the $\Delta\delta\mu_0$ differences are quite small and first
overtone pulsators should give smaller $\Delta\delta\mu_0$ values
than fundamental pulsators with the same period. However the
selection of Cepheids with $P\le$ 10 days is more difficult,
in particular in the inner field (see M06). Therefore, we performed 
a new test by removing these short period variables from the sample 
and we found that the new results are almost identical to those listed 
in Table~7 and Table~8.

Eventually, it seems plausible to suspect that either the Za94
oxygen gradient requires a revision, or the galactic location
cannot be used as a reliable metallicity parameter for individual
Cepheids, or a combination of the above effects.

\clearpage
\begin{figure}
\begin{center}
\includegraphics[width=8cm]{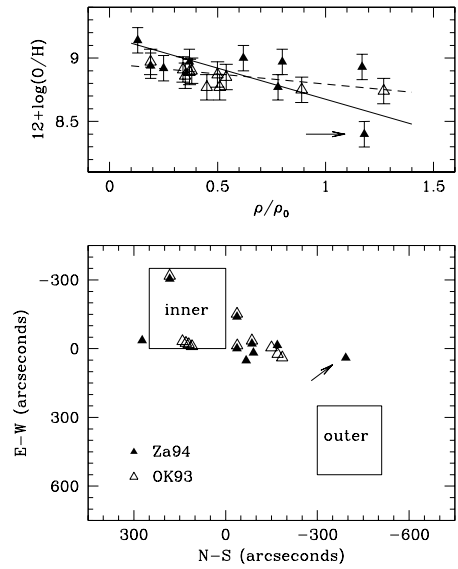}
\caption{Top -- Nebular oxygen abundances measured by Za94 (filled triangles)
and by OK93 (open triangles) versus the fractional isophotal radius with
$\rho_0$=7'.92. The solid line shows the Za94 relation, while the dashed
line is based on equation (16), and it was estimated by neglecting the
H~II region marked by the arrow. Bottom -- Positions of the H~II regions
observed by Za94 and by OK93 in comparison with the NGC~4258 inner and
outer fields.}
\end{center}
\end{figure}
\clearpage

\section{Cepheid metal content and galactic abundance gradient}

In the top panel of Fig.~10, we plot Za94 and previous oxygen abundance
measurements by Oey \& Kennicutt (1993, hereinafter [OK93]) versus
the fractional isophotal radius with $\rho_0$=7'.92. It is
quite clear that the variations in abundance among external
regions with $\rho/\rho_0\sim$1.2 are greater than the abundance
uncertainties and that the lowest value (12+log(O/H)=8.4) measured by
Za94 at $\rho/\rho_0$=1.18 strongly affects the slope of equation (15). 
Indeed, if we neglect this value the linear fit to all the Za94 and 
the OK93 measurements yields (dashed line) a significantly flatter 
gradient, namely

\begin{equation}12+\log(O/H)=8.89-0.16(\rho/\rho_0-0.4)\end{equation}

The bottom panel of Fig.~10 shows that all the Cepheids in the
NGC~4258 inner field are located close to H~II regions where, as
described by equation (16), the oxygen abundance has the almost
constant solar value, i.e. 12+log(O/H)=8.86$\pm$0.08. On the other
hand, the variables in the outer field are distant from any observed
H~II region, and only marginally close to the H~II region underabundant 
in oxygen (12+log(O/H)=8.4). Even though, we assume a tight star-by-star
correlation between oxygen abundance and radial distance, 
we find that by using the equation (16)
to estimate the individual abundances of NGC~4258 Cepheids would
yield a mean abundance difference of only $\sim$ 0.15 dex between
the inner and the outer field. This would imply that the
NGC~4258 is not the right laboratory to constrain the metallicity
effect on the LMC-relative distance moduli. On the other hand, the assumption
of a lower oxygen abundance
for the outer field variables would imply that the galactic gradient 
becomes significantly steeper when moving from the western to the 
eastern direction.

\clearpage
\begin{figure}
\begin{center}
\includegraphics[width=8cm]{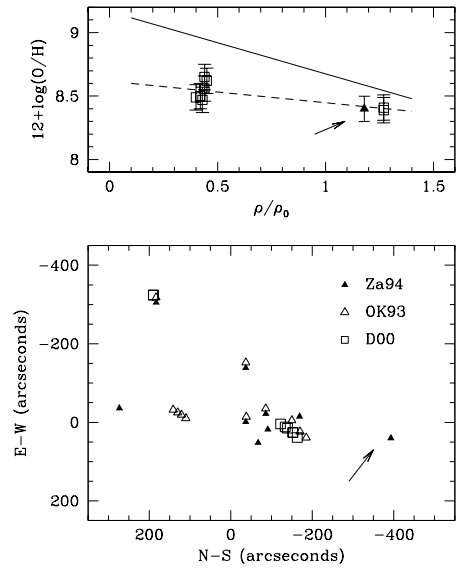}
\caption{Top -- Nebular oxygen abundances measured by D00 versus the
fractional isophotal radius with $\rho_0$=7'.92. The solid line is the
Za94 oxygen gradient, while the dashed line is the best fit line
given by equation (17).
Bottom -- Positions of the H~II regions observed by Za94, by OK93
and by D00.
}
\end{center}
\end{figure}
\clearpage

Although the occurrence of spatial asymmetric metallicity gradients
cannot be ruled out (see, e.g., Kennicutt \& Garnett 1996), we draw
the attention on recent observations specifically addressed to study
extragalactic H~II regions which were expected to be metal-rich.
As a whole, the new measurements (see, e.g., Kennicutt, Bresolin \& Garnett 2003;
Bresolin, Garnett \& Kennicutt 2004; Bresolin et al. 2005) yield a
significant decrease in the
nebular oxygen abundances of regions more metal-rich than LMC, and
marginally affect the abundances of metal-poor ones. Therefore, the
galactic gradients become significantly shallower than those estimated
by previous determinations.

In this context it is worth mentioning that Diaz et al. (2000, hereinafter 
[D00]), by performing a more detailed analysis of optical and near-infrared
observations of several NGC~4258 regions previously observed by
Za94 and by OK93, measured oxygen abundances that are on average a
factor of two lower. Data plotted in Fig.~11 disclose that, by using the
new and more accurate D00 abundances, the NGC~4258 abundance gradient
can be (dashed line)

\begin{equation}12+\log(O/H)=8.55-0.17(\rho/\rho_0-0.4),\end{equation}
\noindent which implies a LMC-like mean oxygen abundance for both
the inner (sample A: [O/H]=$-0.32\pm$0.08) and the outer field
Cepheids (sample B: [O/H]=$-0.49\pm$0.09). This would also imply a
reasonable agreement with the predicted correlation between the
metal abundance and the $\Delta\delta\mu_0$ values.

\section{Conclusions and final remarks}

In the above sections, we have shown that both the comparison
with pulsation models and the most recent H~II abundance measurements
suggest a rather constant, LMC-like metal content for the Cepheids
observed in the two fields of NGC~4258. This finding, once confirmed,
would prevent any reliable differential determination of the $P$-$L$
metallicity dependence. As a consequence, the observed difference
of $\sim$ 0.20 mag in distance modulus between outer and inner field
variables might be caused by other observational effects rather than
a difference in metal abundance.

\clearpage
\begin{figure}
\begin{center}
\includegraphics[width=8cm]{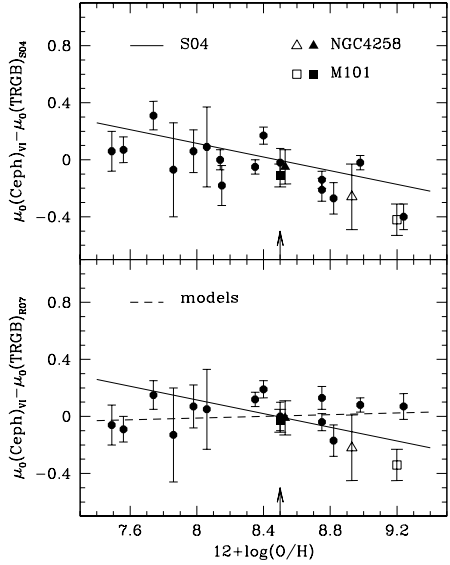}
\caption{Top -- Difference between Cepheid ($VI$-based)
and TRGB distances from Sakai et al. (2004, [S04]), as a function
of the nebular oxygen abundance in the Za94 scale. The arrow
marks the LMC, while filled and open symbols display the outer
and the inner fields in M~101 and in NGC~4258, respectively.
The solid line was drawn using the S04 value $\gamma=-$0.25~mag~dex$^{-1}$.
Bottom -- Same as the top, but with the TRGB distances according to
Rizzi et al. (2007, [R07]). The dashed line was drawn using the value
$\gamma$=+0.05~mag~dex$^{-1}$ of the predicted $P$-$WVI$ relation.}
\end{center}
\end{figure}
\clearpage

We are facing the evidence that the NGC~4258 results presented by
M06 agree quite well with the metallicity effect $\gamma=-$0.24~mag~dex$^{-1}$
determined by Kennicutt et al. (1998) from Cepheid observations in
two fields of M~101. However, it is worthy mentioning that Macri et al. (2001)
brought forward the hypothesis that blended Cepheids could be responsible for
a large fraction of the difference in distance modulus between the outer and
the inner field in M~101. We recall that blended Cepheids, which are mainly
expected in the crowded inner galactic fields, appear brighter than they
really are and that their distances are systematically underestimated by 
$\sim$ 6-9\% (see Mochejska et al. 2000), leading to $\mu_0$ underestimated
by approximately 0.1-0.2 mag.

Moreover, the $\gamma=-$0.25~mag~dex$^{-1}$
provided by Sakai et al. (2004, hereinafter [S04]) from the comparison 
of distances based on Cepheid variables and on the tip of the red giant 
branch (TRGB) has been recently questioned by Rizzi et al. (2007, 
hereinafter [R07]). 
By adopting the distance determinations listed in Table 9, we plot in 
the top panel of Fig.~12 the S04 
difference between the Cepheid, based on the LMC $P$-$L_V$ and
$P$-$L_I$ relations, and the TRGB distances versus the Za94 nebular
oxygen abundances. The data\footnote{Cepheid and TRGB distance scales
are normalized to $\mu_0$(LMC)=18.50 mag. To the S04 original
distances we added the current Cepheid distances to NGC~4258 and
to SMC, and the WLM Cepheid distance by Pietrzynski et al. (2007).}
clearly indicate a trend with metallicity, with the Cepheid residual
distance modulus decreasing with increasing oxygen abundance
and leading to $\gamma=-$0.25~mag~dex$^{-1}$ (solid line).
However, data plotted in the bottom panel of this figure show that 
distance determinations provided by R07 using the TRGB method agree 
within the errors, with Cepheid distances, {\it with the exception 
of M~101 and NGC~4258 inner fields}, over the entire metallicity 
range. Note that in this case we neglected the metallicity correction. 
The agreement becomes even better if we adopt the $\gamma$=+0.05~mag~dex$^{-1}$ 
value of the predicted $P$-$WVI$ relation (dashed line).

\clearpage
\begin{figure}
\begin{center}
\includegraphics[width=8cm]{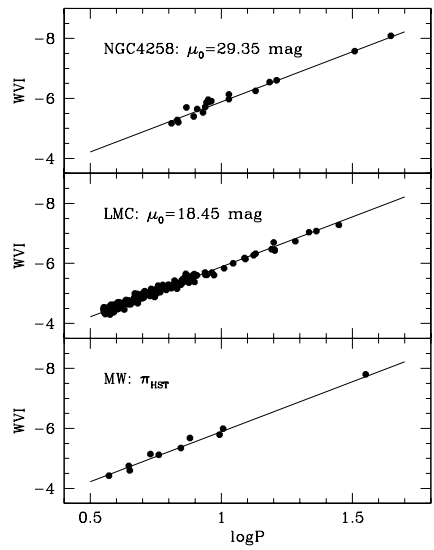}
\caption{Bottom -- Absolute $P$-$WVI$ relation for Galactic Cepheids with
$HST$ trigonometric parallaxes. This relation is used to derive the true
distance modulus of Cepheids in the LMC (middle) and in the outer field
of NGC~4258 (top), by neglecting the metallicity correction.}
\end{center}
\end{figure}
\clearpage

As a final test of the metallicity effect on Cepheid distances based
on $VI$ magnitudes we adopted the Galactic Cepheids with $HST$
trigonometric parallaxes (Benedict et al. 2007). From the absolute
$WVI$ functions of these variables, we find that they obey to the
$P$-$WVI$ relation  WVI=$-2.55-3.33$log$P$, as shown by the solid
line in the bottom panel of Fig.~13.  By using this relation for the
Cepheids in the LMC and in the outer field of NGC~4258 {\it by
neglecting the metallicity correction}, we derive
$\mu_{0,VI}$(LMC)=18.45$\pm$0.09 mag and
$\mu_{0,VI}$(NGC4258)=29.35$\pm$0.12 mag, which are both only
slightly {\it larger} than $\mu_0$(LMC)=18.41$\pm$0.09 mag
determined by Guinan, Ribas \& Fitzpatrick (2004) from eclipsing
binaries (EB)  and $\mu_0$(NGC4258)=29.29$\pm$0.15 mag based on the
maser geometric distance measured by Herrnstein et al. (1999). But,
the variables in the LMC and in the outer field of NGC~4258 {\it do}
have a lower metal abundance by $\sim-0.4$ dex than the Galactic
variables and the adoption of the M06 value
$\gamma=-$0.29~mag~dex$^{-1}$ would {\it increase} the distance
moduli to $\mu_{0,VI}$(LMC)=18.56$\pm$0.09 mag and
$\mu_{0,VI}$(NGC4258)=29.46$\pm$0.12 mag, causing metallicity
corrected Cepheid distances which are larger by about 0.15-0.17 mag
than EB and maser distance determinations. It goes without saying
that by adopting $\gamma$=+0.05~mag~dex$^{-1}$ from the predicted
$P$-$WVI$ relation would further improve the consistency between the
Cepheid distances and the quoted EB and maser-based determinations.

In summary, the main findings of the current paper are the following:

the theoretical pulsation models suggest that both the sign
and the amount of the metallicity dependence of the $P$-$W$
relations depend on the chosen passbands. In particular, for
distances based on $BVI$ magnitudes, the predicted metallicity
effect on $\mu_0$ varies from $\gamma\sim-$0.61~mag~dex$^{-1}$
($P$-$WBV$ relation) to $\gamma\sim$+0.05~mag~dex$^{-1}$ ($P$-$WVI$
relation) over the range $Z$=0.004-0.04. These predictions are
supported by the comparison of SMC and Milky Way Cepheids with LMC
variables.

Accurate $BVI$ photometry of Cepheids in two fields of
NGC~4258 leads to a systematic difference in the true distance
moduli of $\sim$+0.2 mag between the outer and the inner field.
Adopting for individual Cepheids the oxygen abundance given by
their galactocentric distance and the abundance gradient of Z94,
one derives a metallicity effect $\gamma\sim-$0.29~mag~dex$^{-1}$
which is consistent with an earlier
$\gamma\sim-$0.24~mag~dex$^{-1}$ found by Kennicutt et al. (1998)
from Cepheids in two fields of M~101.

The comparison with pulsation models as well as with Magellanic
and Galactic variables, indicates a rather small abundance difference
between the NGC~4258 inner and outer fields, in agreement with recent
nebular oxygen abundances by Diaz et al. (2000).

As a whole, the two ``direct'' determinations of the metallicity
effect which provide negative metallicity dependence $\gamma\sim-$0.24 and
$-$0.29~mag~dex$^{-1}$ appear undermined by the lack of a significant 
difference in metal abundance (NGC~4258) or by the possible occurrence
of blended Cepheids in the inner field (M~101).

The comparison of $VI$-based Cepheid distances with independent
determinations based on the TRGB (external galaxies), $HST$
trigonometric parallaxes (Milky Way Cepheids), eclipsing binaries
(Large Magellanic Cloud) and water maser (NGC~4258) does not support
the negative empirical $\gamma$ values. Current results
seem to favor the predicted value $\gamma\sim$+0.05~mag~dex$^{-1}$.

\acknowledgements
It is a pleasure to thank A. Diaz and L. Rizzi for useful discussions
on nebular abundance measurements and on distance determinations to external 
galaxies based on the TRGB method. We also thank N. Patat for several insights on
the projected distances, F. Thevenin and A. Walker for a detailed reading of 
an early draft of this paper. We acknowledge an anonymous referee for his/her 
suggestions that improved the content and the readability of the manuscript.



\clearpage
\tablewidth{0pt}
\begin{deluxetable}{llll}
\scriptsize
\tablecaption{Basic parameters of fundamental pulsation models. The 
adopted luminosity refers to Mass-Luminosity relations based on 
canonical (``can'') evolutionary computations or deals with higher 
luminosity levels (``over'') produced by mild convective core 
overshooting and/or mass loss before or during the Cepheid phase.}
\tablehead{
\colhead{$Z$}&
\colhead{$Y$}&
\colhead{$M/M_{\odot}$}&
\colhead{$\log L/L_{\odot}$}\\
\colhead{(1)}&
\colhead{(2)}&
\colhead{(3)}&
\colhead{(4)}}
\startdata
 0.004&    0.25&    3.5-11.0&      can, over    \\
 0.008 &   0.25&    3.5-11.0&     can, over \\
 0.01  &   0.26&    5.0-11.0 &     can      \\
 0.02  &   0.25, 0.26, 0.28, 0.31& 5.0-11.0  & can, over \\
 0.03  &   0.275, 0,31, 0,335 &  5.0-11.0  &  can       \\
0.04  &   0.25, 0.29, 0.33 &  5.0-11.0  &   can   \\
\enddata
\end{deluxetable}

\tablewidth{0pt}
\begin{deluxetable}{lllll}
\scriptsize
\tablecaption{Predicted $P$-$W$ relations for fundamental pulsators with
$Z$=0.004-0.04 and based on intensity-averaged magnitudes.}
\tablehead{
\colhead{$W$}&
\colhead{$\alpha$}&
\colhead{$\beta$}&
\colhead{$\gamma$}&
\colhead{$\delta$}\\
\colhead{(1)}&
\colhead{(2)}&
\colhead{(3)}&
\colhead{(4)}&
\colhead{(5)}}
\startdata

\multicolumn{5}{c}{$W$=$\alpha$+$\beta$log$P$+$\gamma$log$(Z/X)$+$\delta$log$(L/L_c)$}\\
$WBV$&$-3.90\pm$0.09&$-3.79\pm$0.03&$-0.61\pm$0.03 &$+0.64\pm$0.04\\
$WVI$&$-2.82\pm$0.13& $-3.24\pm$0.05 & $+0.05\pm$0.03  & $+0.81\pm$0.04\\
$WBI$&$-3.05\pm$0.09& $-3.36\pm$0.02&$-0.11\pm$0.03  &$+0.76\pm$0.04 \\
$WBVI$&$-3.47\pm$0.06& $-3.57\pm$0.02&$-0.35\pm$0.03 &$+0.70\pm$0.04\\
\enddata
\end{deluxetable}

\tablewidth{0pt}
\begin{deluxetable}{lcccccc}
\scriptsize
\tablecaption{Average LMC-relative absolute distance moduli of canonical
($L$=$L_c$) fundamental pulsation models with the labeled metal ($Z$) and helium
($Y$) content.}
\tablehead{
\colhead{$Z$}&
\colhead{$Y$}&
\colhead{$\log (Z/X)$}&
\colhead{$\delta\mu_{0,BV}$}&
\colhead{$\delta\mu_{0,VI}$}&
\colhead{$\delta\mu_{0,BI}$}&
\colhead{$\delta\mu_{0,BVI}$}\\ 
\colhead{(1)}&
\colhead{(2)}&
\colhead{(3)}&
\colhead{(4)}&
\colhead{(5)}&
\colhead{(6)}&
\colhead{(7)}}
\startdata
0.004& 0.250  &$-$2.271&$-$18.46$\pm$0.06&$-$18.75$\pm$0.12&$-$18.66$\pm$0.09&$-$18.57$\pm$0.05\\
0.008& 0.250  &$-$1.967&$-$18.72$\pm$0.05&$-$18.77$\pm$0.10&$-$18.75$\pm$0.08&$-$18.74$\pm$0.05\\
0.010& 0.260  &$-$1.863&$-$18.72$\pm$0.07&$-$18.73$\pm$0.12&$-$18.72$\pm$0.09&$-$18.72$\pm$0.06\\
0.020& 0.250  &$-$1.562&$-$18.99$\pm$0.08&$-$18.70$\pm$0.10&$-$18.76$\pm$0.09&$-$18.88$\pm$0.06\\
0.020& 0.260  &$-$1.556&$-$18.97$\pm$0.07&$-$18.74$\pm$0.11&$-$18.79$\pm$0.09&$-$18.88$\pm$0.05\\
0.020& 0.280  &$-$1.544&$-$18.94$\pm$0.08&$-$18.67$\pm$0.12&$-$18.73$\pm$0.11&$-$18.84$\pm$0.09\\
0.020& 0.310  &$-$1.525&$-$18.90$\pm$0.04&$-$18.65$\pm$0.11&$-$18.70$\pm$0.09&$-$18.80$\pm$0.05\\
0.030& 0.275  &$-$1.365&$-$19.05$\pm$0.07&$-$18.72$\pm$0.09&$-$18.79$\pm$0.08&$-$18.92$\pm$0.06\\
0.030& 0.310  &$-$1.342&$-$19.00$\pm$0.07&$-$18.64$\pm$0.09&$-$18.71$\pm$0.09&$-$18.86$\pm$0.07\\
0.030& 0.335  &$-$1.326&$-$18.98$\pm$0.07&$-$18.64$\pm$0.08&$-$18.71$\pm$0.08&$-$18.85$\pm$0.07\\
0.040& 0.250  &$-$1.249&$-$19.20$\pm$0.08&$-$18.82$\pm$0.07&$-$18.90$\pm$0.06&$-$19.05$\pm$0.06\\
0.040& 0.290  &$-$1.224&$-$19.13$\pm$0.06&$-$18.77$\pm$0.06&$-$18.84$\pm$0.07&$-$18.99$\pm$0.06\\
0.040& 0.330  &$-$1.197&$-$19.07$\pm$0.08&$-$18.71$\pm$0.06&$-$18.78$\pm$0.06&$-$18.93$\pm$0.07\\
\enddata
\end{deluxetable}

\tablewidth{0pt}
\begin{deluxetable}{ccccccc}
\small 
\tablecaption{Internal differences among the LMC-relative absolute
distance moduli listed in Table 1 for fundamental pulsators with the
labeled metal ($Z$) to hydrogen ($X$) ratio and log$P$=0.4-1.9. The
coefficients of the linear least-squares fits to the data are given
in the last two lines.}
\tablehead{
\colhead{$\log (Z/X)$}&
\colhead{$\Delta\delta\mu_{0,BV-VI}$}&
\colhead{$\Delta\delta\mu_{0,BV-BI}$}&
\colhead{$\Delta\delta\mu_{0,BV-BVI}$}&
\colhead{$\Delta\delta\mu_{0,BI-VI}$}&
\colhead{$\Delta\delta\mu_{0,BVI-VI}$}&
\colhead{$\Delta\delta\mu_{0,BVI-BI}$}\\ 
\colhead{(1)}&
\colhead{(2)}&
\colhead{(3)}&
\colhead{(4)}&
\colhead{(5)}&
\colhead{(6)}&
\colhead{(7)}}
\startdata
$-$2.27   &   +0.27$\pm$0.14    &  +0.19$\pm$0.11 &  +0.11$\pm$0.06 &+0.07$\pm$0.05 &+0.16$\pm$0.08 &+0.09$\pm$0.06\\
$-$1.97   &   +0.04$\pm$0.11    &  +0.02$\pm$0.09 &  +0.02$\pm$0.05 &+0.02$\pm$0.04 &+0.02$\pm$0.08 &+0.01$\pm$0.05\\
$-$1.86   &   +0.01$\pm$0.14    &  +0.00$\pm$0.12 &  +0.01$\pm$0.06 &+0.01$\pm$0.05 &+0.01$\pm$0.09 &  $-$0.01$\pm$0.06\\
$-$1.56   &   $-$0.29$\pm$0.13  &$-$0.23$\pm$0.11 &$-$0.11$\pm$0.05 &$-$0.06$\pm$0.05 &  $-$0.18$\pm$0.09 &  $-$0.11$\pm$0.05\\
$-$1.55   &   $-$0.23$\pm$0.15  &$-$0.18$\pm$0.12 &$-$0.09$\pm$0.06 &  $-$0.05$\pm$0.05 &  $-$0.14$\pm$0.09 &  $-$0.09$\pm$0.06\\
$-$1.54   &   $-$0.29$\pm$0.10  &$-$0.23$\pm$0.08 &$-$0.11$\pm$0.04 &  $-$0.06$\pm$0.03 &  $-$0.17$\pm$0.06 &  $-$0.11$\pm$0.04\\
$-$1.53   &   $-$0.25$\pm$0.11  &$-$0.20$\pm$0.09 &$-$0.10$\pm$0.05 &  $-$0.05$\pm$0.05 &  $-$0.15$\pm$0.07 &  $-$0.10$\pm$0.05\\
$-$1.36   &   $-$0.34$\pm$0.08  &$-$0.27$\pm$0.07 &$-$0.13$\pm$0.04 &  $-$0.07$\pm$0.04 &  $-$0.20$\pm$0.06 &  $-$0.13$\pm$0.03\\
$-$1.34   &   $-$0.36$\pm$0.07  &$-$0.28$\pm$0.06 &$-$0.14$\pm$0.03 &  $-$0.07$\pm$0.03 &  $-$0.22$\pm$0.05 &  $-$0.14$\pm$0.03\\
$-$1.33   &   $-$0.34$\pm$0.06  &$-$0.27$\pm$0.06 &$-$0.13$\pm$0.03 &  $-$0.07$\pm$0.03 &  $-$0.21$\pm$0.05 &  $-$0.14$\pm$0.03\\
$-$1.25   &   $-$0.37$\pm$0.10  &$-$0.30$\pm$0.08 &$-$0.15$\pm$0.04 &  $-$0.08$\pm$0.03 &  $-$0.23$\pm$0.06 &  $-$0.15$\pm$0.04\\
$-$1.22   &   $-$0.36$\pm$0.05  &$-$0.28$\pm$0.04 &$-$0.14$\pm$0.03 &  $-$0.07$\pm$0.02 &  $-$0.22$\pm$0.03 &  $-$0.14$\pm$0.03\\
$-$1.20   &   $-$0.36$\pm$0.06  &$-$0.29$\pm$0.05 &$-$0.14$\pm$0.03 &  $-$0.07$\pm$0.03 &  $-$0.22$\pm$0.04 &  $-$0.15$\pm$0.02\\

\multicolumn{7}{c}{$\Delta\delta\mu_0$=$A$+$B$log$(Z/X$)}\\

$A$  &   $-$1.15$\pm$0.05&$-$0.89$\pm$0.04&$-$0.45$\pm$0.02 &$-$0.26$\pm$0.02&$-$0.70$\pm$0.03&$-$0.44$\pm$0.02\\
$B$   &   $-$0.60$\pm$0.05&$-$0.46$\pm$0.04&$-$0.24$\pm$0.02&$-$0.14$\pm$0.02&$-$0.37$\pm$0.03&$-$0.23$\pm$0.02\\
\enddata
\end{deluxetable}

\clearpage 

\tablewidth{0pt}
\begin{deluxetable}{lccccc}
\small 
\tablecaption{Average LMC-relative absolute distance moduli for LMC and
SMC fundamental Cepheids with $P\ge$ 6 days. The last line gives the
corrected values for SMC variables according to the predicted
metallicity effects: $\gamma(\delta\mu_{0,BV})\sim-$0.59~mag~dex$^{-1}$,
$\gamma(\delta\mu_{0,VI}\sim$+0.11~mag~dex$^{-1}$,
$\gamma(\delta\mu_{0,BI})\sim-$0.12~mag~dex$^{-1}$, and
$\gamma(\delta\mu_{0,BVI})\sim-$0.35~mag~dex$^{-1}$}
\tablehead{
\colhead{Galaxy}&
\colhead{$[Fe/H]$}&
\colhead{$\delta\mu_{0,BV}$}&
\colhead{$\delta\mu_{0,VI}$}&
\colhead{$\delta\mu_{0,BI}$}&
\colhead{$\delta\mu_{0,BVI}$}\\ 
\colhead{(1)}&
\colhead{(2)}&
\colhead{(3)}&
\colhead{(4)}&
\colhead{(5)}&
\colhead{(6)}}
\startdata

LMC          &$-$0.35 &+0.01$\pm$0.15 &$-$0.01$\pm$0.08    &+0.00$\pm$0.08 &+0.00$\pm$0.10\\
SMC          &$-$0.70 &+0.78$\pm$0.20&+0.49$\pm$0.14&+0.58$\pm$0.14&+0.68$\pm$0.16\\
SMC$^1_{cor}$&$-$0.70 &$\sim$+0.57   &$\sim$+0.53   &$\sim$+0.54   &$\sim$+0.56\\
\enddata
\end{deluxetable}

\tablewidth{0pt}
\begin{deluxetable}{lcccccc}
\small 
\tablecaption{Differences among LMC-relative absolute distance moduli for
LMC and SMC fundamental Cepheids with $P\ge$ 6 days. The last two
lines give the observed variations between SMC and LMC variables and
the predicted values for $\Delta$[Fe/H]=$-$0.35 according to the
$B$-values listed in Table 4.}
\tablehead{
\colhead{Galaxy}&
\colhead{$\Delta\mu_{0,BV-VI}$}&
\colhead{$\Delta\mu_{0,BV-BI}$}&
\colhead{$\Delta\mu_{0,BV-BVI}$}&
\colhead{$\Delta\mu_{0,BI-VI}$}&
\colhead{$\Delta\mu_{0,BVI-VI}$}&
\colhead{$\Delta\mu_{0,BVI-BI}$}\\ 
\colhead{(1)}&
\colhead{(2)}&
\colhead{(3)}&
\colhead{(4)}&
\colhead{(5)}&
\colhead{(6)}&
\colhead{(7)}}
\startdata
LMC&  +0.03$\pm$0.15&+0.01$\pm$0.11&+0.01$\pm$0.06&+0.02$\pm$0.04&+0.02$\pm$0.09&+0.00$\pm$0.05\\
SMC&  +0.33$\pm$0.19&+0.23$\pm$0.14&+0.13$\pm$0.08&+0.09$\pm$0.05&+0.20$\pm$0.12&+0.10$\pm$0.07\\

SMC$-$LMC & $\sim$+0.30&$\sim$+0.22&$\sim$+0.12&$\sim$+0.07&$\sim$+0.18&$\sim$+0.10\\
$\Delta$[Fe/H]=$-$0.35&  $\sim$+0.21&$\sim$+0.16&$\sim$+0.08&$\sim$+0.05&$\sim$+0.13&$\sim$+0.08\\
\enddata
\end{deluxetable}
\tablewidth{0pt}
\begin{deluxetable}{lcccccc}
\small 
\tablecaption{Average LMC-relative absolute distance moduli for NGC~4258 
Cepheids with $\rho/\rho_0<$0.7 (sample A) and $\rho/\rho_0>$1.0 (sample B).}
\tablehead{
\colhead{Sample}&
\colhead{$\langle\rho/\rho_0\rangle$}&
\colhead{$[O/H]$\tablenotemark{a}}&
\colhead{$\delta\mu_{0,BV}$}&
\colhead{$\delta\mu_{0,VI}$}&
\colhead{$\delta\mu_{0,BI}$}&
\colhead{$\delta\mu_{0,BVI}$}\\ 
\colhead{(1)}&
\colhead{(2)}&
\colhead{(3)}&
\colhead{(4)}&
\colhead{(5)}&
\colhead{(6)}&
\colhead{(7)}}
\startdata

A &0.40$\pm$0.22 &+0.13$\pm$0.08&10.69$\pm$0.25 &10.62$\pm$0.23&10.65$\pm$0.20 &10.67$\pm$0.20\\
B &1.40$\pm$0.24 &$-$0.37$\pm$0.09&10.94$\pm$0.25 &10.85$\pm$0.12&10.88$\pm$0.12 &10.90$\pm$0.17\\

\enddata
\tablenotetext{a}{Based on the Za94 oxygen gradient.} 
\end{deluxetable}

\tablewidth{0pt}
\begin{deluxetable}{lccccccc}
\tabletypesize{\footnotesize} 
\tablecaption{Averaged differences among LMC-relative absolute distance
moduli for NGC~4258 Cepheids with $\rho/\rho_0<$0.7 (sample A) and 
$\rho/\rho_0<$1.0 (sample B), in comparison with Magellanic and 
metal-rich Milky Way (MW) variables.}
\tablehead{
\colhead{Sample}&
\colhead{$[O/H]$}&
\colhead{$\Delta\delta\mu_{0,BV-VI}$}&
\colhead{$\Delta\delta\mu_{0,BV-BI}$}&
\colhead{$\Delta\delta\mu_{0,BV-BVI}$}&
\colhead{$\Delta\delta\mu_{0,BI-VI}$}&
\colhead{$\Delta\delta\mu_{0,BVI-VI}$}&
\colhead{$\Delta\delta\mu_{0,BVI-BI}$}\\ 
\colhead{(1)}&
\colhead{(2)}&
\colhead{(3)}&
\colhead{(4)}&
\colhead{(5)}&
\colhead{(6)}&
\colhead{(7)}&
\colhead{(8)}}
\startdata

A&  +0.13$\pm0.08$\tablenotemark{a}&+0.07$\pm$0.28&+0.04$\pm$0.21&+0.03$\pm$0.11&+0.03$\pm$0.08&+0.04$\pm$0.17&+0.01$\pm$0.10\\
B&$-$0.37$\pm0.09$\tablenotemark{a}&+0.09$\pm$0.24&+0.06$\pm$0.18&+0.04$\pm$0.10&+0.03$\pm$0.07&+0.05$\pm$0.15&+0.02$\pm$0.08\\

SMC&$-$0.88$\pm0.08$\tablenotemark{b}&  +0.31$\pm$0.17&+0.21$\pm$0.14&+0.13$\pm$0.07&+0.09$\pm$0.05&+0.19$\pm$0.11&+0.10$\pm$0.07\\
LMC&$-$0.37$\pm0.15$\tablenotemark{b}&  +0.03$\pm$0.15&+0.01$\pm$0.11&+0.01$\pm$0.06&+0.02$\pm$0.04&+0.02$\pm$0.09&+0.00$\pm$0.05\\
MW &+0.15$\pm0.06$\tablenotemark{c}&$-$0.13$\pm$0.12&$-$0.12$\pm$0.10&$-$0.05$\pm$0.05 &$-$0.01$\pm$0.04&$-$0.07$\pm$0.08&$-$0.06$\pm$0.05\\

\enddata
\tablenotetext{a}{Za94 oxygen gradient.} 
\tablenotetext{b}{Ferrarese et al. (2000)} 
\tablenotetext{c}{Galactic Cepheids with [Fe/H]$_A$=0.1-0.3, and by adopting $[O/H]=[Fe/H]$} 
\end{deluxetable}

\clearpage 
\tablewidth{0pt}
\begin{deluxetable}{lclll}
\small 
\tablecaption{TRGB and Cepheid distances determined by Sakai et al. (2004, 
 [S04]) and by Rizzi et al. (2007, [R07]).}
\tablehead{
\colhead{Galaxy}&
\colhead{$12+log(O/H)$}&
\colhead{$\mu_0$(TRGB)}&
\colhead{$\mu_{0,VI}$(Cep)}&
\colhead{$\mu_0$(TRGB)}\\ 
\colhead{}&
\colhead{(Za94)}&
\colhead{(R07)}&
\colhead{(S04)}&
\colhead{(S04)}}
\startdata

LMC        &  8.50    &   18.59$\pm0.09$  &   18.50$\pm0.10$  &   18.57$\pm0.06$ \\
SMC        &  7.98    &   18.99$\pm0.08$  &   18.99$\pm0.15^1$    &   18.98$\pm0.06$ \\
IC1613     &  7.86    &   24.31$\pm0.06$  &   24.17$\pm0.33$  &   24.38$\pm0.05$ \\
IC4182     &  8.40    &   28.25$\pm0.06$  &   28.35$\pm0.06$  &   28.23$\pm0.05$ \\
NGC224     &  8.98    &   24.47$\pm0.11$  &   24.38$\pm0.05$  &   24.37$\pm0.08$ \\
NGC300     &  8.35    &   26.65$\pm0.07$  &   26.53$\pm0.05$  &   26.48$\pm0.04$ \\
NGC598     &  8.82    &   24.81$\pm0.04$  &   24.47$\pm0.11$  &   24.71$\pm0.06$ \\
NGC3109    &  8.06    &   25.52$\pm0.05$  &   25.54$\pm0.28$  &   25.57$\pm0.05$ \\
NGC3351    &  9.24    &   30.39$\pm0.13$  &   29.92$\pm0.09$  &   29.92$\pm0.05$ \\
NGC3621    &  8.75    &   29.36$\pm0.11$  &   29.15$\pm0.06$  &   29.26$\pm0.12$ \\
NGC3031    &  8.75    &   28.03$\pm0.12$  &   27.75$\pm0.08$  &   27.70$\pm0.04$ \\
NGC4258$_i$&  8.93    &   29.46$\pm0.11$  &   29.12$\pm0.23$\tablenotemark{a}    &   29.42$\pm0.06$\tablenotemark{b} \\
NGC4258$_o$&  8.53    &   29.46$\pm0.11$  &   29.35$\pm0.12$\tablenotemark{a}    &   29.42$\pm0.06$\tablenotemark{b} \\
NGC5253    &  8.15    &   27.88$\pm0.11$  &   27.63$\pm0.14$  &   99.00$\pm0.00$ \\
NGC5457$_i$&  9.20    &   29.42$\pm0.11$  &   28.93$\pm0.11$  &   29.34$\pm0.09$ \\
NGC5457$_o$&  8.50    &   29.42$\pm0.11$  &   29.24$\pm0.08$  &   29.34$\pm0.09$ \\
NGC6822    &  8.14    &   23.37$\pm0.07$  &   23.30$\pm0.07$  &   99.00$\pm0.00$\\
SexA       &  7.49    &   25.67$\pm0.13$  &   25.66$\pm0.14$  &   25.78$\pm0.06$\\
SexB       &  7.56    &   25.63$\pm0.04$  &   25.63$\pm0.09$  &   25.79$\pm0.04$\\
WLM        &  7.74    &   24.77$\pm0.09$  &   25.01$\pm0.10$\tablenotemark{c}    &   24.93$\pm0.04$\\

\enddata
\tablenotetext{a}{This paper.} 
\tablenotetext{b}{Macri et al. (2006).} 
\tablenotetext{c}{Pietrzynski et al. (2007).} 
\end{deluxetable}

\end{document}